# Enriching meta-analyses through scoping review, bibliometrics, and alternative impact metrics: Visualizing study characteristics, hidden risk of bias, societal influence, and research translation

Yefeng Yang[a,*], Malgorzata Lagisz[a,#], Shinichi Nakagawa[a,#]

[a] Evolution & Ecology Research Centre and School of Biological, Earth and Environmental Sciences, University of New South Wales, Sydney, NSW 2052, Australia

[#] Equally contributed senior author

* Corresponding author

yefeng.yang1@unsw.edu.au (Y.Y)

**Open Science**

Raw data and analytical script to reproduce examples presented in the manuscript are archived at GitHub repository https://github.com/Yefeng0920/MA_Map_Bib.

**Abstract**

We present a framework consisting of three approaches that can enhance meta-analyses: 1) scoping reviews (evidence map), 2) bibliometrics, and 3) alternative impact metrics. These three 'enrichment' approaches facilitate the research synthesis of both quantitative and qualitative evidence, along with academic and non-academic influences. While the meta-analysis yields quantitative insights (e.g., overall estimates), the enrichment analyses provide user-friendly summaries of qualitative information on the evidence base. Scoping reviews can visualize study characteristics, unravelling knowledge gaps and methodological differences. Bibliometric analysis offers a visual assessment of the non-independent evidence, such as hyper-dominant authors and countries, and funding sources, potentially informing the risk of bias. Impact metric analysis employs alternative metrics to gauge societal influence and research translation (e.g., policy and patent citations) of studies in the meta-analysis. To illustrate the application of this framework, we provide sample visualizations and R code.

# 1 | INTRODUCTION

Data visualization is an effective tool for knowledge communication. It also works in meta-analysis [1-3], a statistical method for quantitatively synthesizing evidence [4-12]. The incorporation of data visualization can yield two potential benefits to meta-analysis. First, data visualisation can facilitate the identification of patterns, trends, and associations, not immediately apparent within the complex meta-analytic data [13]. Second, it could increase the uptake of meta-analytic findings by the end-users in science, policy, and practice [2,14,15].

Typical meta-analytic data includes both quantitative and qualitative information. Conventionally, the meta-analytic visualization methods have primarily focused on the quantitative aspects [1,13], exemplified by forest plots and funnel plots for representing effect sizes and their precisions [3]. However, qualitative information tends to be conveyed in a format not readily digestible by end-users [16]. Yet, these qualitative details are no less valuable than their quantitative counterparts. Notably, qualitative information comes in two types: study characteristics (e.g., contextual factors of the performed research) and bibliographic attributes (e.g., author and country information).

A user-friendly summary of study characteristics can add value to meta-analyses. For example, visualizing the intersections between the types of studies interventions and outcomes (or other combinations of two of categorical variables) can identify knowledge gaps [17], thus setting the agenda for future research to fill such gaps. At the same time, identifying evidence "gluts" is of urgent importance given research waste due to preventable problems [18], such as new research ignoring the existing evidence [19]. The need for visualization aligns with the tenets of scoping reviews, sometimes termed evidence review or map (see [20] [21] for a debate) [22,23]. Interestingly, network meta-analyses offer a precursor to integrating evidence mapping into meta-analyses. The geometry of network graphs can visually unveil the state of the evidence, distinguishing well-studied interventions from those warranting further investigation [24].

Similarly, visualising bibliographic attributes remains relatively rare in meta-analyses. Yet, such visualizations can reveal the non-independence of evidence [25]. For example, bibliometric analyses can map author collaboration networks [26], defining research groups and detecting hyper-dominant author clusters (see [27] for an example) [28]. Such hyper-dominance induces non-independent evidence and could influence between-study level Risk of Bias (RoB). Bibliometric analysis of traditional impact metrics (e.g., citations) can be used to map the conceptual influences within academia. Additionally, alternative impact metrics [29-32], such as Altmetric scores, patents, and policy citations, can be mapped to reflect the influences outside academia.

In this work, we provide a framework that enriches traditional meta-analyses by seamlessly integrating three approaches: evidence mapping, bibliometric and alternative impact analyses (Figure 1; https://yefeng0920.github.io/MA_Map_Bib/). Also, we use data from diverse disciplines to illustrate the principles, values, and procedures, as examples in Boxes. The elegance of this framework lies in converting the qualitative information into readily digestible formats for end-users with minimal additional data extraction effort from meta-analysts. The proposed framework can also enrich other research synthesis methods in the systematic-review family (e.g., systematic reviews without meta-analysis), review-of-review family (e.g., overview, umbrella review), and rapid-review family (e.g., rapid review) [33,34].

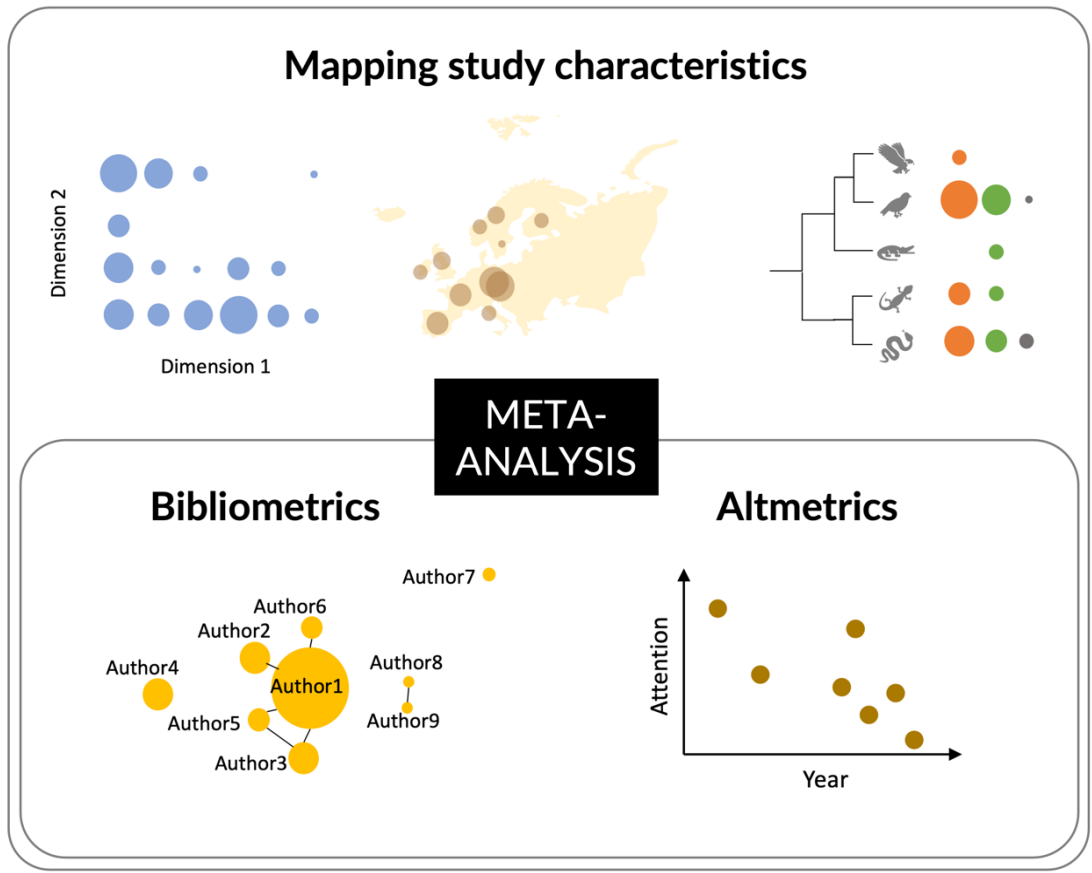

**Figure 1.** A conceptual diagram of implementing enrichment plots in meta-analyses. Selected examples for the types of plots discussed in the main text.

## 2 | FRAMEWORK FOR INTEGRATING SCOPING REVIEW, BIBLIOMETRICS, AND ALTMETRICS INTO META-ANALYSIS

The three enrichment analyses of qualitative information (details see below) can be implemented in conjunction with a broad spectrum of meta-analytic approaches used across different fields, including conventional random-effects, multilevel, multivariate, dose-response, longitudinal, network, location-scale, individual participant data (IPD), phylogenetic comparative models, and meta-meta-analysis (see Box 1 – 3) [4,11]. Below, we introduce the typical applications of the three enrichment approaches.

### 2.1 | Evidence mapping within meta-analyses

Scoping reviews represent a form of research synthesis approach that systematically collates and catalogues existing evidence on a specific topic, yet is broader than a systematic review [35]. While similar approaches exist with varying terminologies, such as mapping reviews, systematic evidence maps, and evidence and gap maps [20-22,24,36,37], here we treat them as synonyms because they share a common core feature: mapping the state of the evidence, which makes them valuable for meta-analysis enrichment. We use the term "evidence mapping", defining it as a process that employs various visualisation methods to provide a user-friendly summary of study characteristics (qualitative evidence) within a meta-analysis (the end-product; Box 1) [16,21]. Incorporating evidence mapping into a meta-analysis offers two benefits. Firstly, it inherits the merits of a conventional evidence map, such as identifying knowledge gaps, directing research priorities, and providing support for funding and policy decisions [16,24]. Secondly, evidence mapping can inform methodological decisions (e.g., *post hoc* analyses) and interpretations of meta-analytic findings. Various visualization approaches exist, such as heatmap and bubble plots [38] and other displays used in the 3ie Evidence Gap Map [16] and Evidence-based Policing Matrix [39], for mapping the evidence landscape.

We recommend using grid-like graphs, where one dimension (e.g., intervention) is placed on the x-axis, and the other (e.g., outcomes) is placed on the y-axis, with statistics (e.g., the number of studies, number of effects, sample size) displayed at the intersection of the x- and y-axes (Figure I in Box 1) [17]. Additional dimensions, such as study design, effect size, study quality or RoB (Risk of Bias) measures, can be also incorporated (see more on this point in the next section). We recommend using Sankey or alluvial diagrams to visualize the flow or overlaps in the composition of context-dependence drivers (Figure II in Box 1), summarizing their connections and co-linearity and missing data patterns in an accessible manner. Comprehensive and clear visual representation of moderator variables can facilitate customizable evidence synthesis identifying context-specific drivers and delivering more tailored evidence for science, policy, and practice, improving generalizability and transferability of evidence (see details in Box 1) [40]. Additionally, a geographical map of study participants could reveal the representation status of historically underrepresented regions or groups, by incorporating location, gender, ethnic or social backgrounds into visualisations. Understanding such patterns, in turn, can facilitate actions towards greater equity, diversity, and inclusion (EDI) in research. For phylogenetic meta-analytic analyses, visual representation of species information, such as species phylogenetic trees (Figure III in Box 1), can intuitively convey the breadth of taxa and underlying phylogenetic heterogeneity.

## 2.2 | Bibliometric analysis within meta-analyses

Bibliometric analysis, often called bibliometrics, is a technique that effectively synthesizes bibliometric data to provide an overview of the performance and intellectual structure of a field [41]. It holds a unique position in the research synthesis method ecosystem because it synthesizes the bibliometric characteristics (metadata of the literature, not the content inside it) [25], such as authors, countries, funders, and citation counts [42]. Literature-level characteristics is to scientific papers as epidemiology is to patients [43]. Diagnosing the "epidemiological susceptibility risk" of studies can uncover hidden biases across evidence base. Such a process can also extend the scope of RoB assessment from within-study to between-study levels (note that between-study RoB is usually known as publication bias while within-study RoB is just 'risk of bias'). Studies are often not independent in terms of their conduct, with same researchers potentially involved in multiple studies. Thus, highly represented authors potentially dominating the production of evidence, limiting methodological diversity and generalizability. Indeed, centralized collaboration communities, which involve shared authors and employ similar protocols, have been shown to yield less replicable findings than decentralized ones, which incorporate more diverse independent research groups [44]. Moreover, the publication of false, exaggerated, and falsified effects is believed to be more common in countries with a "publish or perish" culture [45,46]. For example, meta-scientific evidence suggests that studies from the United States tend to overestimate effect sizes and exhibit a larger publication bias [47-49].

Bibliometrics can bring two major additions to meta-analyses. Bibliometrics can construct a co-authorship network for the studies included in a meta-analysis (Figure IV in Box 2). Mapping the co-authorship network plot can intuitively reveal authorship dependence [28]. Network metrics, such as the degree of centrality, can provide quantitative insights. Bibliometrics can also identify dominant and unrepresented countries of author affiliation, revealing interdependences between countries (Figure V in Box 2). Following similar principles, bibliometrics can identify research location bias (e.g., studies from high-impact or predatory journals) [50,51], funding source bias [52,53], linguistic bias [54,55], and time-lag bias [56,57]. Notably, a quantitative assessment of those factors can be accomplished by conducting subgroup analysis or meta-regression [4].

## 2.3 | Alternative impact analysis within meta-analyses

Analysis of alternative impact metrics, termed as "altmetrics", is an approach used to estimate research impact beyond academia [32]. It can quantify societal impact by tracking activities on social media (e.g., Facebook and X), Internet pages (e.g., Wikipedia), policy-related documents, and patent applications [32,58-60]. Incorporating altmetrics into a meta-analysis can reveal the uptake of research findings by stakeholders outside academia. For example, funding agencies have explored to utilize altmetrics to assess public engagement with the research they support [61-63]. Similarly, policymakers have used it to determine early engagement with research within the policy domain [64]. We recommend visualising the overall scores (e.g., Altmetric score), policy citations, and patent citations to identify the overall social media attention received by studies within a meta-analysis and the extent to which these studies are translated into practical applications, respectively (Box 3).

## 3 | FINAL REMARKS

Meta-analysis has become a widely accepted synthesis method across various disciplines. Meanwhile, scoping reviews (evidence maps), bibliometrics, and altmetrics are gaining popularity. We propose integrating these three approaches with meta-analysis. For meta-analysts, this framework requires minimal effort to locate, screen, code, and appraise additional evidence, as these procedures are typically integral to a standard meta-analysis. For end-users (e.g., researchers, policymakers, and practitioners), it summarizes the qualitative part of meta-analytic evidence base in immediately digestible graphs enriched with additional dimensions. These graphs can better visualize the landscape of the meta-analytic evidence base, facilitate the identification of knowledge gaps and methodological decisions, uncover hidden biases, and map the societal impact and degree of research translation.

**BOX 1: Mapping meta-analytic evidence bases**

We demonstrate the application of evidence mapping using three publicly available meta-analytic datasets from various disciplines [65-67]. Note that these applications are provided for illustrative purposes, and any scientific, clinical, or policy implications should not be drawn from them. The data and code for replicating all visualizations can be found in the https://yefeng0920.github.io/MA_Map_Bib/.

**Data**

*Data 1*: Hodkinson and colleagues [65] conducted a network meta-analysis to assess the efficacy of different self-management interventions (multidisciplinary case management, regularly supported self-management, and minimally supported self-management) in enhancing the quality of life among asthma patients.

*Data 2*: Mertens and colleagues [66] employed a multilevel meta-analytic model to synthesize evidence on the effectiveness of choice architecture interventions (often referred to as nudges) for behaviour change across various techniques, behavioural domains, and other study characteristics (e.g., populations and locations).

*Data 3*: Sanders and colleagues [67] used a Bayesian meta-analytic model to synthesize evidence regarding the impacts of artificial light at night on physiological, phenological, life history, activity patterns, and population/community-based outcomes. This meta-analysis included more than 180 species. For illustration, we used the subset that focused on physiological outcomes.

**Methods**

Evidence maps (scoping reviews) often use structured formats to map the evidence of a given field, such as PCC (Participant, Context, Concept) and UTOS (Units, Treatments, Outcomes, or Settings) [17,20]. For Data 1, we used grid-like graphs with the intervention variable on the x-axis and the outcome variable on the y-axis [17]. This aimed to identify knowledge gaps in self-management interventions. Each cell displayed two statistics: the number of studies and the number of effect sizes. We further quantified population mean effect size estimate for each cell using a robust point and variance estimation approach [68]. For Data 2, we utilized a Sankey diagram to illustrate the connections (collinearities) between moderator variables related to hypotheses tested. Data 3 featured a phylogenetic tree to visualize the taxonomic breadth and relatedness among the species involved. Data processing and visualizations were conducted using packages *ggplot2* [69], *ggsankey* [70], *ape* [71], *ggtree* [72], and *metafor* [73].

**Results**

*Data 1*: Figure I shows the landscape of evidence about the effectiveness of the self-management interventions. For example, panels A – B visually represent where randomized controlled trials (RCTs) have been conducted to examine the interventions' effectiveness, offering insights into which interventions have received more clinical attention. Panel C reveals a demographic bias, as most self-management interventions were trialled on adults. Panel D conveys critical information about both which interventions have been examined in the RCTs and their associated effectiveness. This information can inform funding strategies (e.g., funding missing RCTs, or in-depth investigation) and help clinicians to gauge the volume and effectiveness of their interventions of interest.

*Data 2*: Figure II highlights the diversity of experimental designs in the primary studies included, suggesting potential heterogeneity in the meta-analytic evidence. The tested moderator variables display minimal collinearity, indicating that each variable represents a unique contextual influence. Importantly, Figure II provides useful visual clues to identify the contexts requested by decision-makers, facilitating the assessment of the effectiveness of interventions in the context of interest (e.g., target population and location). A follow-up customizable evidence synthesis can be conducted to improve the generalizability and transferability of meta-analytic evidence (see [40] for more details).

*Data 3*: Figure III presents a typical phylogenetic tree revealing the broad coverage of taxa used in artificial light at night experiments, including birds, mammals, insect, reptiles, and arachnids. For a more in-depth statistical analysis, constructing a phylogenetic correlation matrix can quantify the effect of the shared evolutionary history among species in the meta-analytic evidence base.

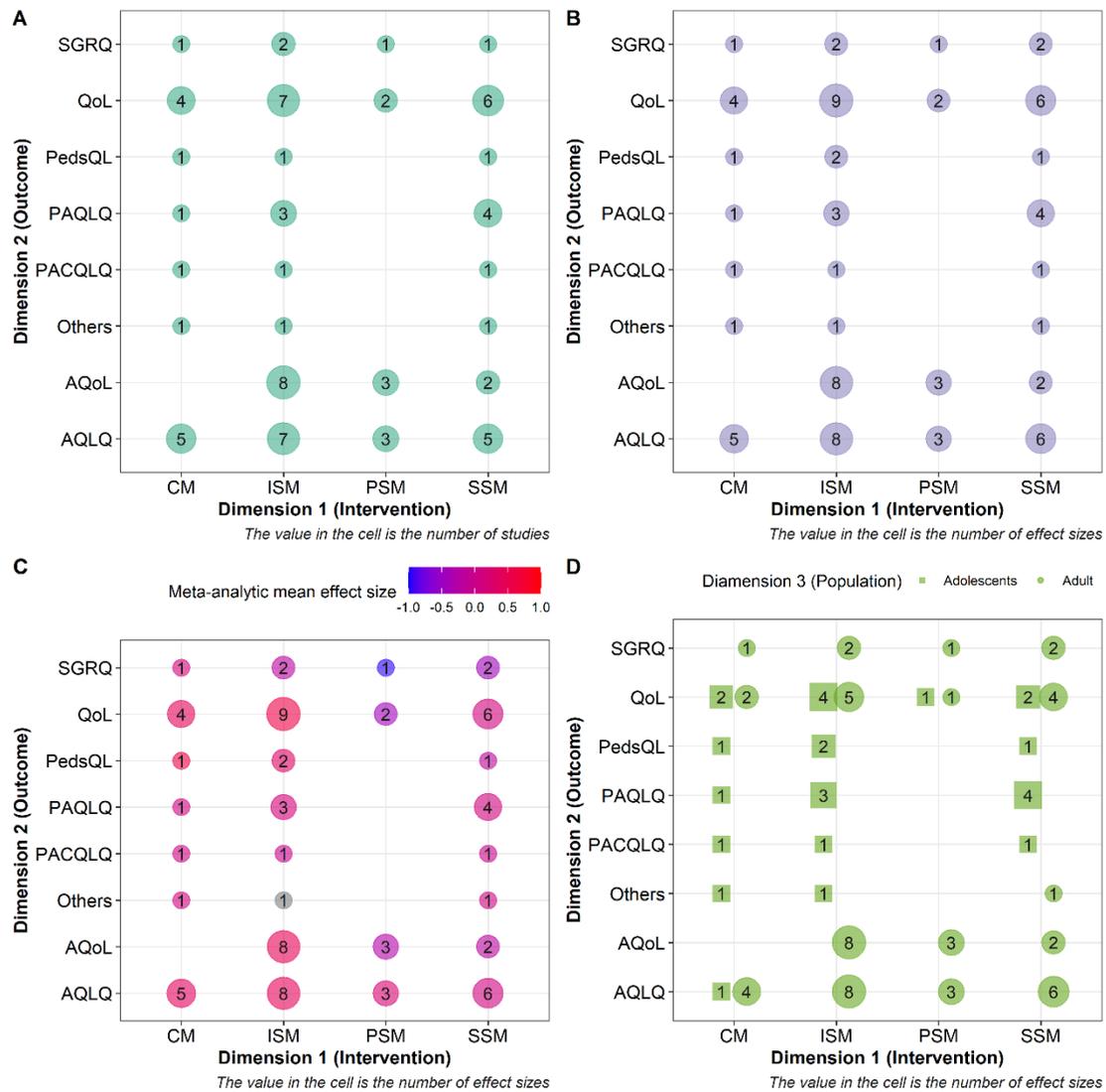

**Figure I** Examples of evidence maps visualizing study characteristics. (A) A typical grid-like graph with intervention variable as the first dimension, outcome variable as the second dimension and the bubble size representing the number of studies. (B) The bubble sizes are changed to represent the number of effect sizes. (C) The colour scale is applied to the bubbles to denote the magnitude of the mean effect size. (D) The population variable is mapped to the shape serving as the third information dimension. R code was adapted from Polanin et al. [17]

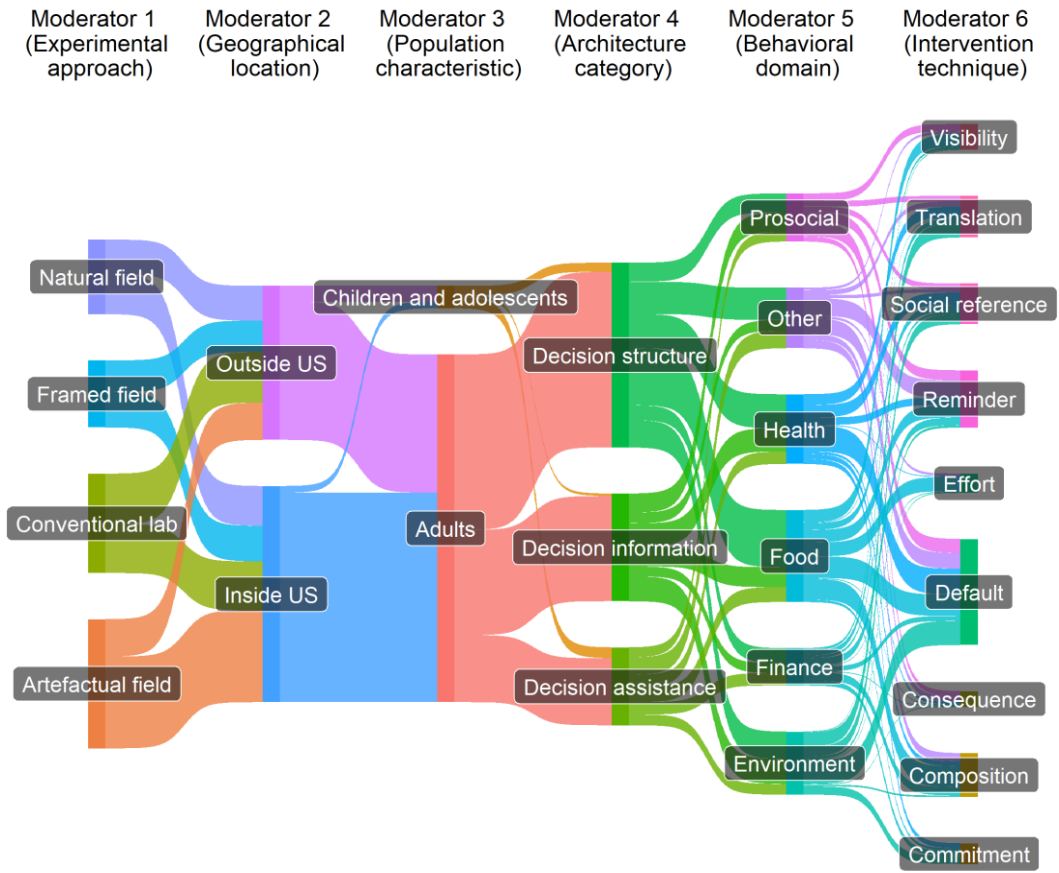

**Figure II** An example of a Sankey diagram showing the flow and change in the composition of moderator variables considered as context-dependence drivers.

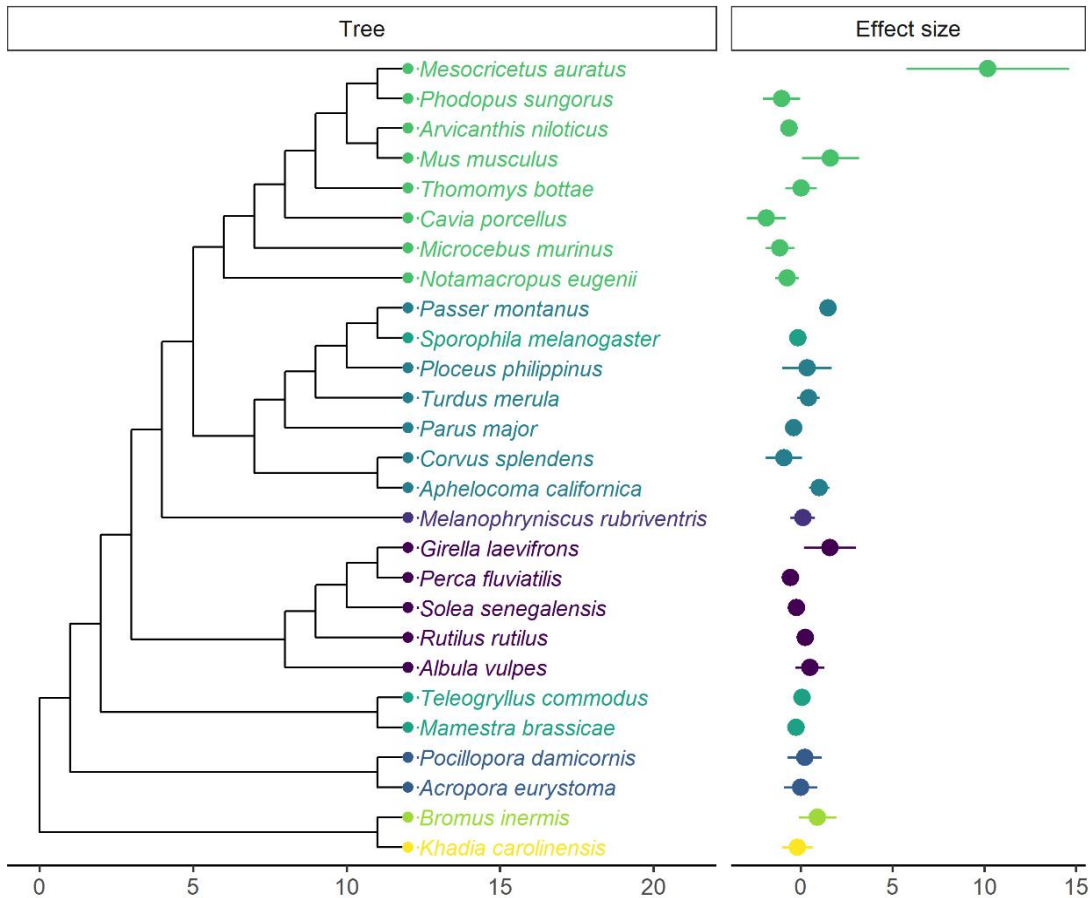

**Figure III** An example of a phylogenetic tree visualizing the breadth of taxa and underlying phylogenetic heterogeneity. The effect size estimate for each species was aggregated from multiple estimates within the same species, assuming a constant within-species correlation (in this case, 0.5). Different colours represent different phylogenetic classes.

**BOX 2: Bibliometric mapping meta-analytic evidence base**

We use Data 3 from Box 1 [67] to illustrate the application of bibliometric analysis in meta-analytic evidence base. The data and code for replicating all visualizations can be found in the https://yefeng0920.github.io/MA_Map_Bib/.

**Data**

A brief description can be found in Data 3 in Box 1 [67].

**Methods**

We collected bibliographic data of the primary studies included in the meta-analysis from Scopus using DOI-based searches. We created two bibliographic network typologies using the associated data: co-authorship and country networks. The co-authorship network was constructed through collaboration analysis by multiplying two matrices, *Author × Paper* and *Paper × Author*. For the country network, bibliographic coupling was used, involving multiplication of *Paper × Cited paper* and *Cited paper × Paper*. We used *bibliometrix* package [26] to construct network. Packages *igraph* [74] and *circlize* [75] were used to project the networks.

**Results**

Figure IV visually depicts the co-author network, with vertices (circles) representing authors and edges (lines) representing co-authorships. Highly-represented authors and dominant research groups can produce the majority of the evidence, reducing its methodological diversity and generalisability (certain groups might report certain results that are caused by bias in experimental designs and analyses) [28]. In the example meta-analytic data set, we observe nearly 90 author clusters, each of which, on average, included six authors. The largest cluster had 58 authors. To quantitatively detect the effect of hyper-dominant research groups on meta-analytic estimates, a leave-one-out analysis can be employed [4] to compare the meta-analytic effect size estimates of each research group with that of the rest. Figure V shows the country of affiliation citation network with 34 countries of author affiliation. While there is no obvious indication of dominant countries (with country contributions shown as proportion of the circle's perimeter), the United States, the United Kingdom, and Germany are the most prominent players in this field and all countries appear to be linked well via article citations. At the statistical follow-up work, a multilevel meta-analytic model with random effects at the levels of author and country of affiliation clusters can be resorted to correct for potential biases in country influence [4,28].

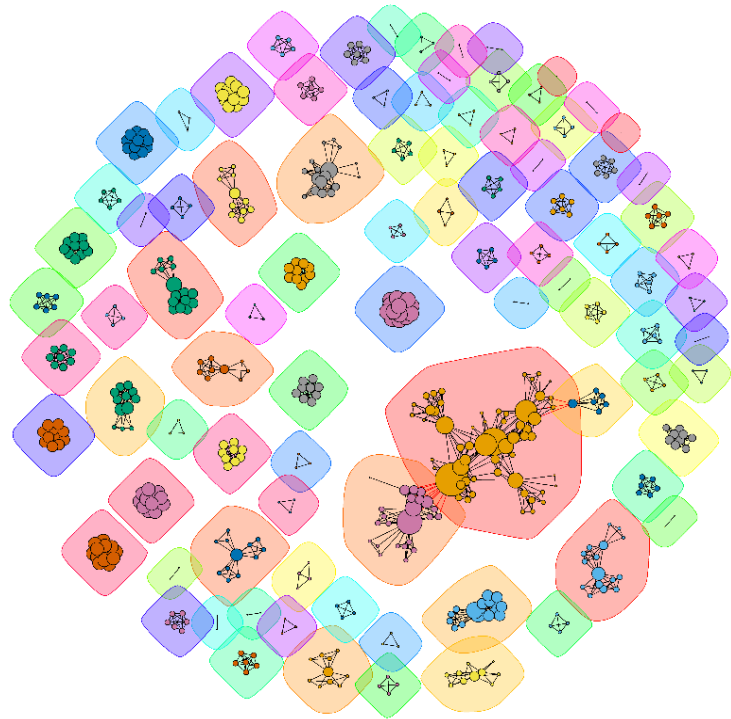

**Figure IV** An example of co-authorship network visualizing the diversity of the research groups and the degree of centralization of the scientific community. The vertices (nodes) and edges (links) denote authors and co-authorships, respectively. The bubbles represent the author clusters. Each colour denotes a co-authorship cluster (or research group). The figure was inspired by Moulin, et al. [28], who originally implemented it using Matlab and VOSviewer.

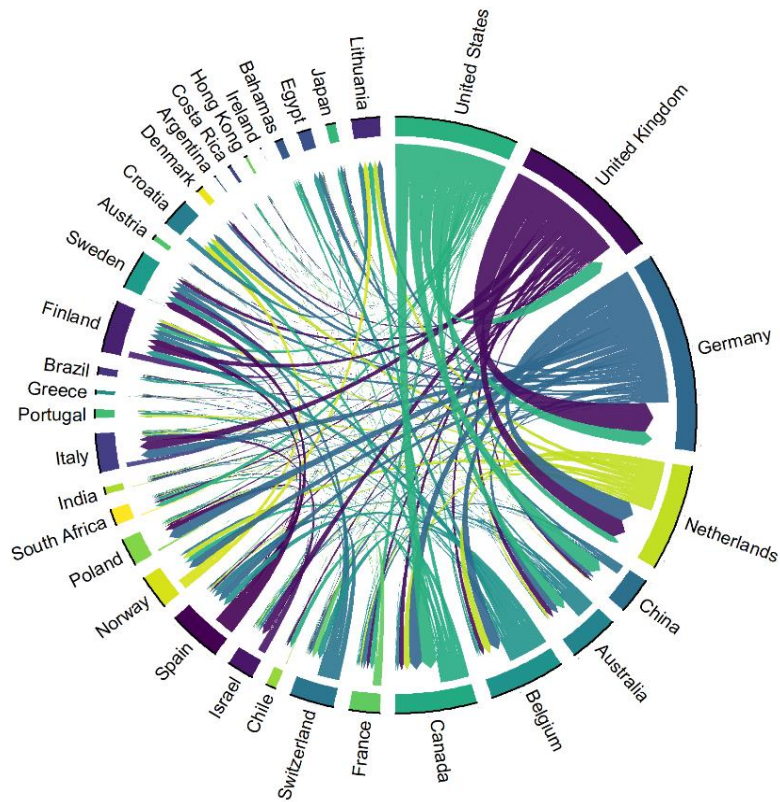

**Figure V** An example of a chord diagram showing the epistemological interdependences between different countries of author affiliation in the meta-analytic evidence base. These interdependences are quantified using a bibliographic coupling approach. Two countries are coupled when the cumulative bibliographies of their respective papers share one or more cited references. The coupling strength, an indicator of the dominance, increases as the number of co-cited references between them increases.

**BOX 3: Alternative metric analysis of meta-analytic evidence base**

We use one meta-analytic data et examining the replicability of the preclinical cancer biology [76] to illustrate the application of almetrics analysis, based on data from Altmetric. All data and code necessary for reproducing the subsequent visualizations can be accessed in the https://yefeng0920.github.io/MA_Map_Bib/.

**Data**

The data used for this demonstration is derived from the work of Errington and colleagues [76], who conducted a meta-analysis on the differences between the original effect size estimates and replicated effect size estimates across 50 cancer biology experiments. These experiments were sourced from 23 papers published in high-profile journals, such as *Nature*, *Science*, *Cell*, and *PNAS*.

**Methods**

We obtained Altmetric scores for each paper included in the meta-analytic dataset using the Altmetrics online service ([www.altemetric.com](www.altemetric.com)). This online service also provides data on the counts of policy documents and patent applications directly mentioned by organizational websites. We used API to automatically retrieve the total Altmetric score, policy, and patent citation for each original publication. We visually represented the results using the orchard plot [77] and grid-like graph [17].

**Results**

Figure VI, panel A, indicates the substantial social media attention earned by cancer biology studies included in the meta-analytic evidence base. A large portion of these studies were mentioned in policy documents and patent applications, indicating a potential degree of practical translation. Notably, full replicated studies exhibited relatively higher Altmetric scores, and larger policy and patent citation counts compared to that of studies that were only partially replicated and not replicated. Among the fully replicated studies, those published in *PNAS*, *Cancer Biology*, and *Cell*, exhibited higher impact metrics than those published in *Nature* and *Science*.

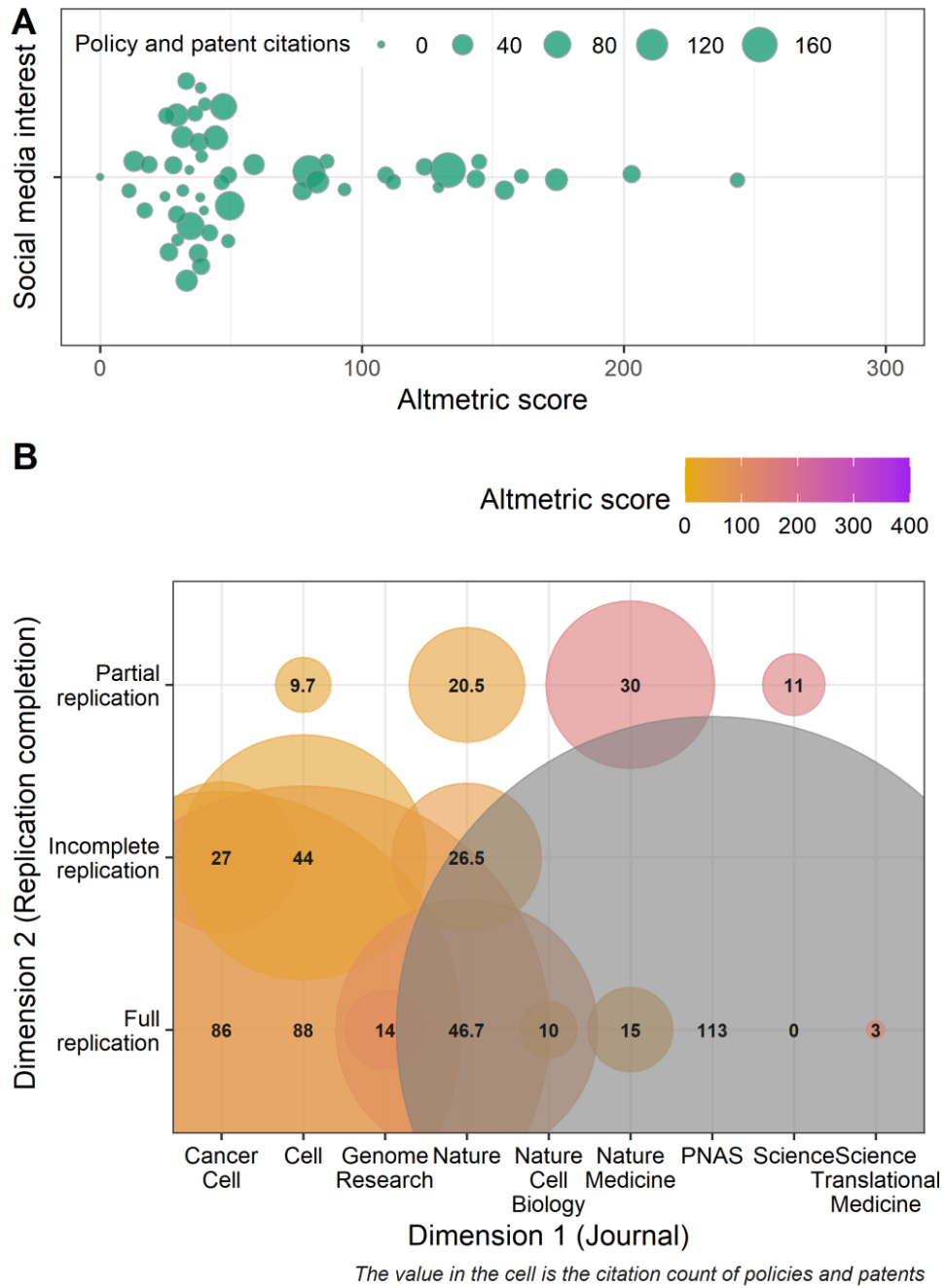

**Figure VI** Examples of visualization showing the results of impact metric analysis. (A) An orchard plot with Altmetric score as the bubble and impact metric related to indicators of practical application (i.e., patent and policy citation counts) as the bubble size. (B) A grid-like graph where: 1) the color and size of the bubbles correspond to the Altmetric score, and 2) the impact metric counts, related to indicator of practical translation (i.e., patent and policy citation counts), respectively. The grey bubble

indicates that the Altmetric score exceeds 400. The categories of "full," "partial," and "no replication" denote that replication of studies was as fully replicated, partially replicated, or not replicated, respectively.

# HIGHLIGHTS

## What is already known

- Data visualization enhances meta-analysis, aiding in identifying patterns and increasing the accessibility of evidence to end-users in science, policy, and practice.
- Quantitative information in meta-analyses is commonly presented through user-friendly formats, such as forest plots and funnel plots, for presenting effect size estimates and precisions.
- Meta-analyses include qualitative information on study characteristics and bibliographic attributes, presented usually in a manner inaccessible by end-users.

## What is new

- Integrating the elements of scoping reviews and bibliometrics into meta-analyses enriches research synthesis by visualising qualitative evidence.
- Scoping reviews (systematic mapping) enhance meta-analyses, providing a user-friendly summary of study characteristics, identifying knowledge gaps, and setting research priorities.
- Bibliometric analysis can visualise non-independent evidence, revealing dominant authors, and countries, informing potential risk of bias.
- Further, alternative impact metrics, such as Altmetric scores, patents, and policy citations, can help capture societal impacts of meta-analyses beyond academia.

## Potential impact for RSM readers outside the authors' field

- The proposed framework transforms qualitative information into easily digestible formats for end-users, requiring minimal additional data extraction effort from meta-analysts.
- The proposed framework's capacity can be extended to other methods in research synthesis ecosystems, such as systematic-review family, review-of-review family, and rapid-review family, to visualise quantitative information for every synthesis.

**FUNDING INFORMATION**

YY was funded by the National Natural Science Foundation of China (NO. 32102597). SN, YY, and ML were funded by the Australian Research Council Discovery Grant (DP210100812 & DP230101248).

**AUTHOR CONTRIBUTIONS**

YY: Conceptualization; data curation; formal analysis; investigation; methodology; software; visualization; writing – original draft; writing – review and editing. ML: Conceptualization; visualization; writing – review and editing; funding acquisition; supervision. SN: Conceptualization; investigation; methodology; software; validation; visualization; writing – review and editing; funding acquisition; supervision. All authors approved the final manuscript.

**CONFLICT OF INTEREST STATEMENT**

The author reported no conflict of interest.

**DATA AVAILABILITY STATEMENT**

The data and analytical script to reproduce examples presented in the manuscript are archived at GitHub reporsitory: https://github.com/Yefeng0920/MA_Map_Bib.

**ORCID**

*Yefeng Yang* 0000-0002-8610-4016

*Malgorzata Lagisz* 0000-0002-3993-6127

*Shinichi Nakagawa* 0000-0002-7765-5182